\documentclass{appolb}

\usepackage{graphicx}

\usepackage{amsmath, amsthm, amssymb}
\usepackage{colortbl}

\begin{document}
\title{Giant CP violation in charmless three-body $B$ meson decays at LHCb: all order formalism for meson-meson final state interactions
\thanks{Presented at ExcitedQCD2024. }%
}
\author{Reyes-Torrecilla, Alba.\footnote{speaker}, Pelaez, Jose R.
\address{Departamento de Física Teórica and IPARCOS, Universidad Complutense de Madrid, E-28040 Madrid, Spain
}
\\[3mm]
{Magalh\~aes, Patricia C.
\address{Departamento de Raios Cósmicos e Cronologia, Universidade Estadual de Campinas 13083-860, Campinas, Brazil}
}
}
\maketitle
\begin{abstract}
LHCb has observed giant CP violation in localized regions of the Dalitz plots of $B$ to three charmless light mesons. This has been interpreted as an enhancement due to strong two-body final state interactions. In this talk, we show how such interactions, described with dispersive analyses of data, can be implemented beyond the leading order expansion in the two-body re-scattering amplitude. 
\end{abstract}
  
\section{Introduction}
\begin{figure}[htb]
\hspace*{-.8cm}
\includegraphics[width=7cm]{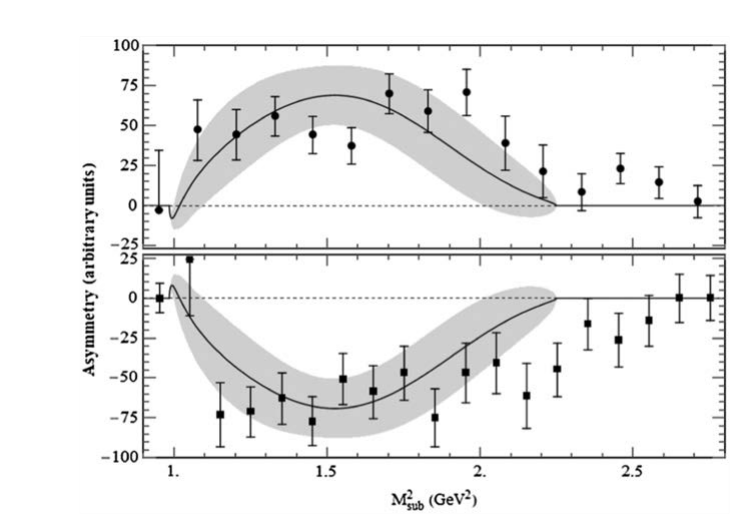} 
\hspace*{-.2cm}
\includegraphics[width=6.4cm, height=4.65cm]{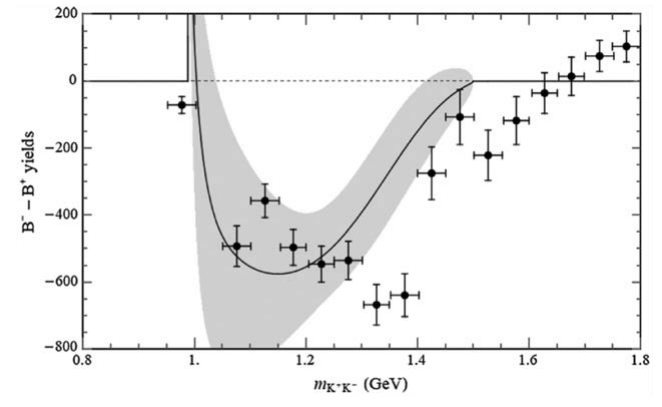}
\hspace*{-.4cm}
\includegraphics[width=6.6cm]{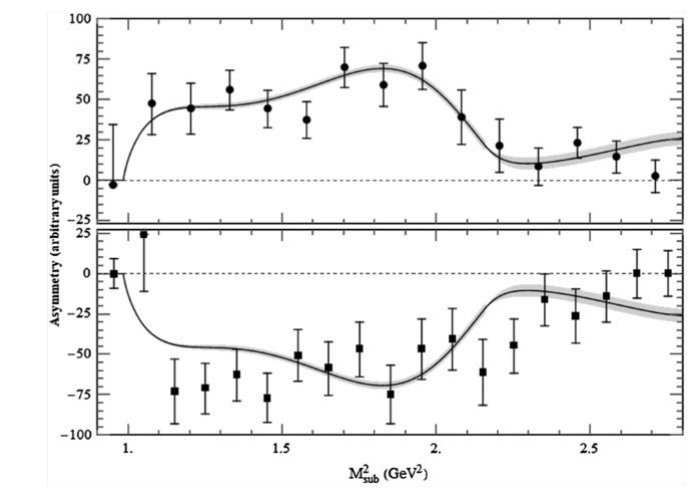} 
\includegraphics[width=6.3cm, height=4.7cm]{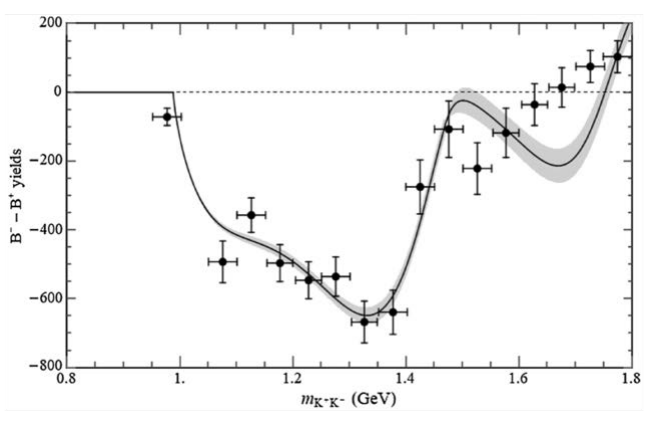}
\hspace*{-.2cm}
\includegraphics[width=6.3cm]{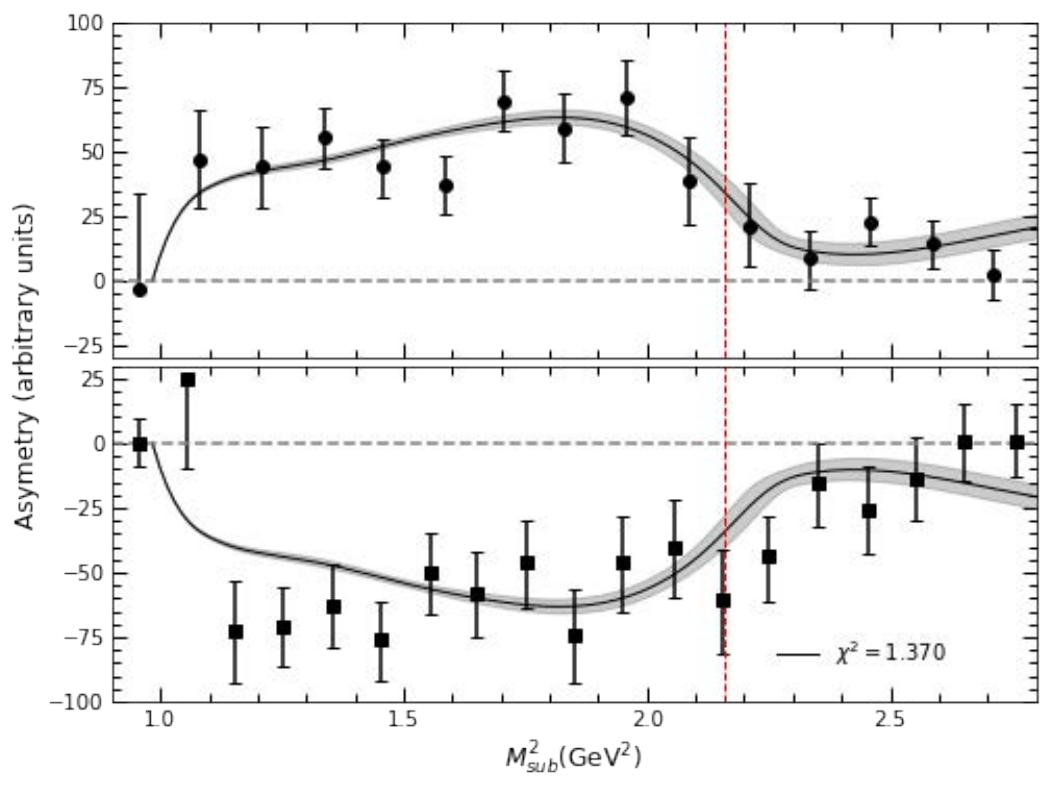} 
\includegraphics[width=6.4cm, height=4.8cm]{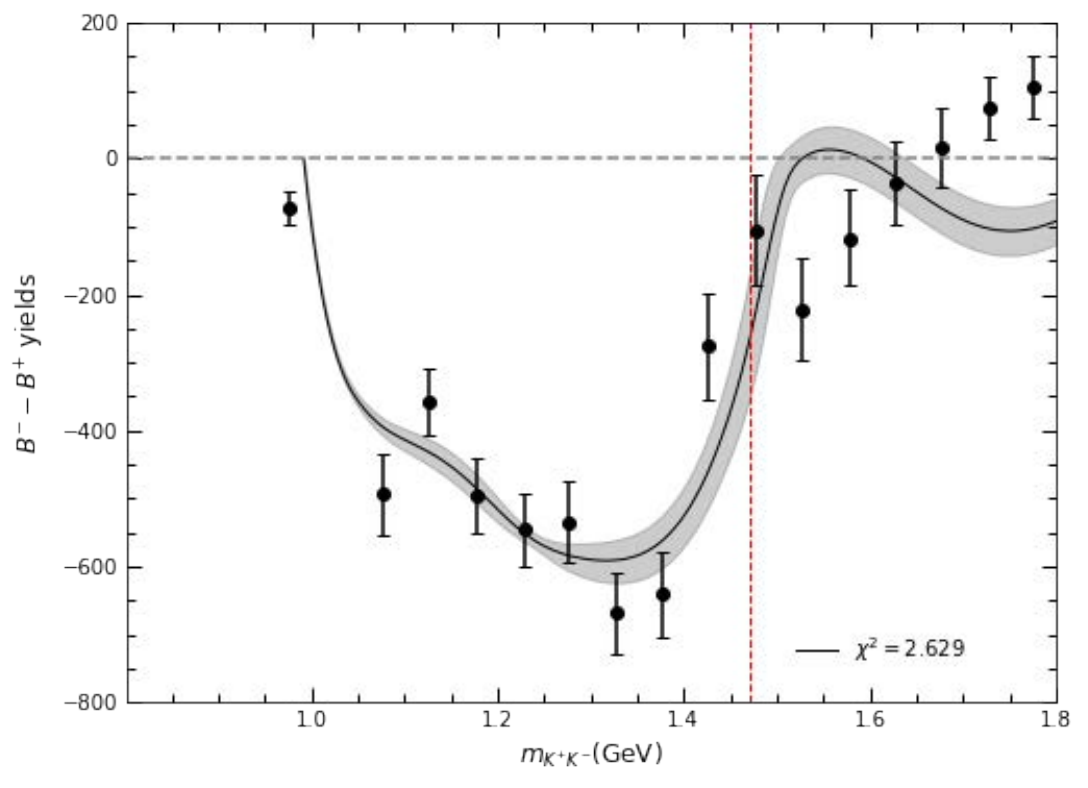}
\caption{Left column: Each pair of panels represents the
CP asymmetries for $B^{\pm}\to K^\pm \pi^+\pi^-$ (up) and 
$B^{\pm}\to K^\pm K^+K^-$ (down). Data from \cite{Aaij:2013sfa}. 
Right panels: Total  $B^{\pm}\to K^\pm K^+K^-$ asymmetry. LHCb data from the sum of Figs 6(c) and (d) in \cite{Aaij:2014iva}. Top panels: Curves using the \cite{Bediaga:2013ela} model with the Eq.\ref{crude} crude estimates. Figures from \cite{Garrote:2022uub}. Central panels: using realistic $\pi\pi\rightarrow K\bar K$ amplitudes from \cite{Garrote:2022uub}. Figures from \cite{Garrote:2022uub}. Low panels: Preliminary ``all-order WS" formalism with realistic $\pi\pi\rightarrow K\bar K$ interactions. 
Error bands come only from the propagation of meson-meson uncertainties. 
}
\label{Fig1}
\end{figure}

Huge CP violations have been observed by LHCb \cite{Aaij:2013sfa,Aaij:2013bla,Aaij:2014iva,LHCb:2022nyw,Aaij:2019qps,Aaij:2019hzr,Aaij:2019jaq}
in the phase space of $B$-meson decays to three light-pseudoscalar meson decays ($M=\pi$, $K$). In particular, for $B^\pm\rightarrow \pi^\pm K^+K^-$, they claim {\em ``the largest CP asymmetry reported to date for a single amplitude of $(-66\pm 4 \pm 2)\%$"}, in the $\pi\pi\rightarrow K\bar K$ S-wave rescattering. Local asymmetries of similar magnitude are seen in other $B\rightarrow 3M$ decays.
The relevance of Final State Interactions (FSI) in this context was already suggested by Wolfenstein and Suzuki \cite{Wolfenstein:1990ks,Suzuki:1999uc,Suzuki:2007je}, although  LHCb analyses \cite{Aaij:2019qps,Aaij:2019hzr,Aaij:2019jaq} used 
the implementation for $B\rightarrow3M$ in \cite{Bediaga:2013ela,Nogueira:2015tsa} (and other models in some cases) at leading order in the rescattering amplitude expansion.  This approach is relevant when two mesons have $1-1.5$ GeV, so that $\pi\pi$ and $K\bar K$ are the dominant channels, and the third meson is a spectator.  CP asymmetries are defined as 
$\Delta\Gamma_\lambda=\Gamma_{B\to\lambda}-\Gamma_{\bar B\to\bar\lambda}$, where $\lambda$, $\lambda'$ label the  two non-spectator mesons.
Then, if CP violation is driven by the  $\pi\pi\rightarrow K\bar K$  isoscalar-scalar (S0) wave FSI, this model predicts opposite CP asymmetries for $B\rightarrow M\pi\pi$ and $B\rightarrow M K\bar K$ decays, which was indeed observed by LHCb
\cite{Aaij:2013sfa} as seen in the pairs of panels of the left column in Fig.\ref{Fig1}. Each pair of panels represents the CP asymmetries for $B^{\pm}\to K^\pm \pi^+\pi^-$ (up) and $B^{\pm}\to K^\pm K^+K^-$ (down). As noted in \cite{Bediaga:2013ela} the fact that one is opposite to the other is a strong indication of the relevant role of FSI.

Unfortunately, the $s$-dependence of these asymmetries was only described qualitatively and with large uncertainties, as seen in the upper panels in Fig.\ref{Fig1}, where  
$M_{sub}$ and $m_{K^+K^-}$ stand for the $\sqrt{s}$ of the interacting meson pair, and
$s$ is the usual Mandelstam variable.  The reason for this qualitative description is that 
the S0 partial wave of  
$\pi\pi\rightarrow K\bar K$  was crudely estimated from that of $\pi\pi\rightarrow\pi\pi$. Namely, for its modulus and phase shift it was assumed 
\begin{equation}
\vert S_{\pi\pi KK}\vert=\sqrt{1-\eta^2}, \quad \delta_{\pi\pi \pi\pi}=\delta_{KKKK}
\Rightarrow
\delta_{\pi\pi KK}=2\delta_{\pi\pi\pi\pi}, \quad 
\label{crude}
\end{equation}
where $\eta(s)$ is the $\pi\pi\rightarrow\pi\pi$ elasticity and
$\delta_i(s)$ are the phase shifts of the corresponding meson-meson scattering channel.
These approximations, however,  do not describe \cite{Garrote:2022uub} the available $\pi\pi\rightarrow K\bar K$ scattering
data.

Nevertheless, it has been very recently shown \cite{Garrote:2022uub} that 
the use of realistic $\pi\pi\rightarrow K\bar K$ amplitudes obtained from a dispersive data analysis \cite{Pelaez:2018qny,Pelaez:2020gnd}, provide a remarkably good description of the such giant CP violation in the 1-1.5 GeV region. As seen in the center panels of Fig.\ref{Fig1} the use
of realistic interactions unveils better the resonant structure and reduces drastically the uncertainty when compared to the panels above. 
The vertical red line stands at 1.47 GeV, up to where the scattering amplitudes were made to satisfy
dispersive constraints in \cite{Pelaez:2018qny,Pelaez:2020gnd}, above that the amplitudes were just simple fits to data.

However, this formalism was applied to leading order in the rescattering partial wave, namely,
\begin{equation}
    {\cal A}^\pm_{LO}=A_\lambda+B_\lambda e^{\pm i\gamma}+i\sum\limits_{\lambda'} t_{\lambda'\lambda} \left(A_\lambda'+B_\lambda' e^{\pm i\gamma} \right).
    \label{eq:simpleALO}
\end{equation}
Here $A_\lambda$, $B_\lambda$ are the CP-symmetric two-meson production amplitudes without FSI, $\gamma$ is the weak phase that changes sign under CP conjugation, and $t_{\lambda\lambda'}$ is the S0 partial wave related to the scattering $S$-matrix partial wave by $S_{\lambda\lambda'}=\delta_{\lambda\lambda'}+2i t_{\lambda\lambda'}$. 

Therefore, given that we are dealing with strong rescattering, it is convenient to study the all-orders ``square root of $S$" Wolfenstein-Suzuki (WS) formalism \cite{Wolfenstein:1990ks,Suzuki:1999uc,Suzuki:2007je}, where:
\begin{equation}
{\cal A}_\lambda= \sum\limits_{\lambda'} S^{1/2}_{\lambda\lambda'} A_\lambda'.
\label{WS}
\end{equation}
Note that the expansion of $S^{1/2}$ to leading order in $t$ yields Eq.\ref{eq:simpleALO}.
This approach had already been followed in \cite{Cheng:2016shb}, but with the crude estimates in Eq.\ref{crude}.
In this talk, we present our preliminary results within the WS formalism and realistic 
meson-meson interactions.

\section{Results}

First of all, in order to calculate the square root of the two-meson $S$-matrix in the S0-wave, we assume that there are just two channels $\pi\pi$ and $K\bar K$. Thus $S$ can be written in terms of two phases and one modulus. A convenient choice is:
\begin{equation}
    S_{\lambda\lambda'}=\begin{pmatrix}
\sqrt{1-\left|S_{\pi\pi\!\!\textup{\tiny{ KK}}}\right|^2}e^{2i\delta_{\pi\pi\pi\pi}} & i\left|S_{\pi\pi\!\!\textup{\tiny{ KK}}}\right|e^{i\delta_{\pi\pi\textup{\tiny{KK}}}} \\
i\left|S_{\pi\pi\!\!\textup{\tiny{ KK}}}\right|e^{i\delta_{\pi\pi\textup{\tiny{KK}}}}& \sqrt{1-\left|S_{\pi\pi\!\!\textup{\tiny{ KK}}}\right|^2}e^{2i\left(\delta_{\pi\pi\textup{\tiny{KK}}}-\delta_{\pi\pi\pi\pi}\right)} 
\end{pmatrix}
\end{equation}
The input for $\vert S_{\pi\pi\!\!\textup{\tiny{ KK}}} \vert$ 
and 
$\delta_{\pi\pi\textup{\tiny{KK}}}$ is
taken from the dispersive data analyses in \cite{Pelaez:2018qny,Pelaez:2020gnd} whereas for $\delta_{\pi\pi\pi\pi}$ is taken from \cite{Pelaez:2019eqa}.

\begin{figure}[htb]
\centering
\includegraphics[width=6.24cm]{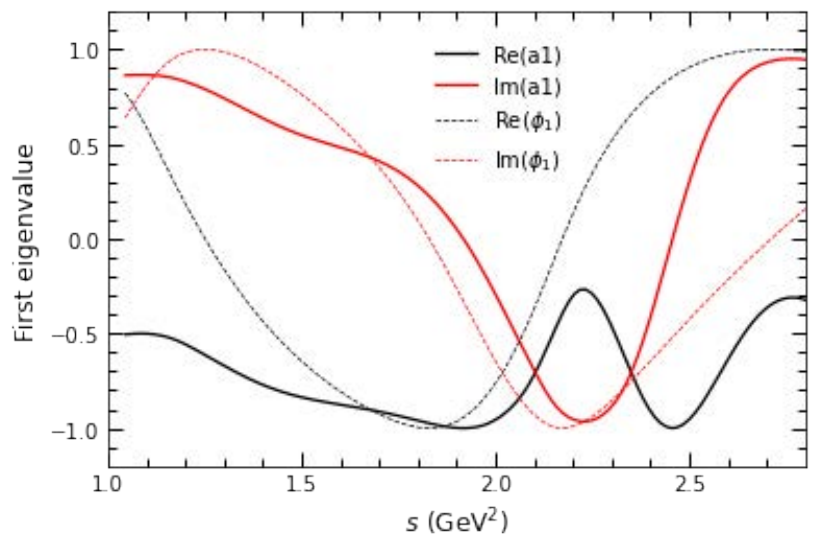} \includegraphics[width=6.24cm]{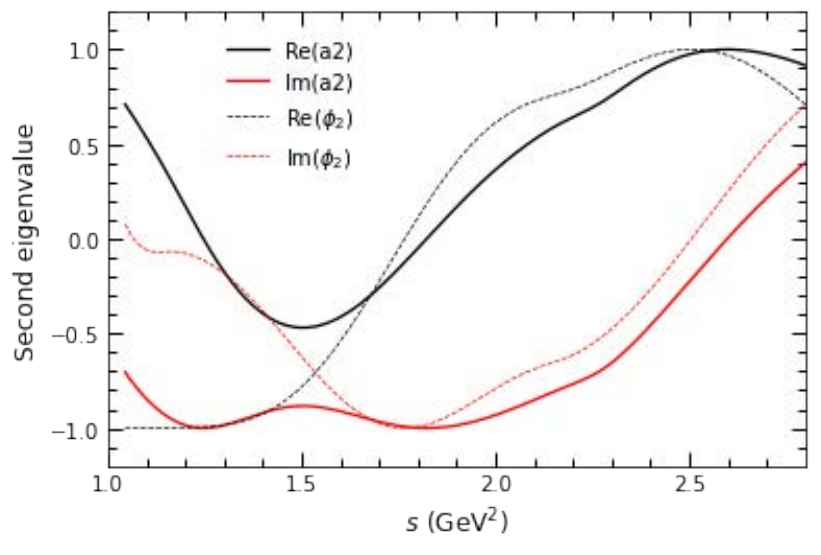} 
   \caption{Continuous lines: Eigenvalues $a_{1,2}$ of the $S0$-partial wave $S$-matrix for two coupled $\pi\pi$ and $K \bar K$ channels, as functions of $s$, obtained from the dispersive data analysis of \cite{Pelaez:2018qny,Pelaez:2020gnd}. Dashed lines: the same eigenvalues, now called $\phi_{1,2}$, when using the crude estimates in Eq.\ref{crude}. }
    \label{Fig2}
\end{figure}

The resulting $s$-dependence of the real and imaginary parts of its two eigenvalues $a_{1,2}$ is shown in Fig.\ref{Fig2}. Note that they are quite different from the eigenvalues (denoted $\phi_{1,2}$) obtained in \cite{Cheng:2016shb}, using the crude estimates of Eq.\ref{crude}.

Once the eigenvalues are calculated it is easy to obtain $S^{1/2}$, 
the ${\cal A}_\lambda$ amplitudes, and then the CP asymmetries.
Our preliminary results are shown in the lower panels of Fig.\ref{Fig2}. 

It can be noticed that the use of the WS formalism also allows for a description of the data while still observing resonant structures. It must be noticed, however, that, although the description is similarly good with the WS formalism, than at LO, the uncertainties are somewhat bigger. This is due to the fact that in the upper and central panel the only FSI contribution that was considered was the one due to $\pi\pi\to K \bar K$ rescattering. In contrast, in the WS formalism we have also included $\pi\pi$ self-interactions, and they add up their own uncertainty.

\section{Summary}
In this talk, we have presented preliminary results, showing that it is possible to extend beyond leading order the existing analysis of the FSI contributions to
describe the data from giant CP violation in charmless three-body decays in the 1-1.5 GeV region. We have applied the 
Wolfenstein-Suzuki ``all-order" formalism, with realistic meson-meson scattering amplitudes as input.

The model is very simple and just describes the dominant FSI contributions, coming predominantly from 
meson-meson interactions in the scalar isoscalar wave, predominantly from
$\pi\pi\to K \bar K$ with some subdominant $\pi\pi\to \pi\pi$ contributions. 
The results are very promising and we are working to extend the model to energies below 1 GeV, as well as to include higher partial wave contributions.

\section*{Acknowledgements}
This work is partially supported by the Grant PID2022-136510NB-C31 funded by MCI /AEI/10.13039/501100011033.
We thank the organizers of the Excited QCD 2024 Workshop for their efforts and 
for creating such a stimulating atmosphere for scientific discussions.

\bibliographystyle{unsrt}
\bibliography{biblio.bib}

\end{document}